\documentstyle[aps,preprint]{revtex}
\tightenlines
\begin{document}
\draft
\preprint{}
\title{Electromagnetic form factors in the $J/\psi$ mass region:
\\The case
in favor of additional resonances.}
\author{N.~N.~Achasov \footnote{e-mail: achasov@math.nsc.ru}
 and A.~A.~Kozhevnikov \footnote{e-mail: kozhev@math.nsc.ru}}
\address{Laboratory of Theoretical Physics,\\
S.L.~Sobolev Institute for Mathematics,\\
630090, Novosibirsk 90, Russian Federation}
\date{\today}
\maketitle
\language=1
\begin{abstract}
Using the results of our recent analysis of $e^+e^-$ annihilation,
we plot the curves for the diagonal and transition form factors
of light hadrons in the time-like region up to the production
threshold of an open charm quantum number. The comparison  with
existing data on the decays of $J/\psi$ into such hadrons shows that some
new resonance structures may be present in the mass range between 2 GeV
and the $J/\psi$ mass. Searching them may help in a better understanding
of the mass spectrum in both the simple and a more sophisticated quark models,
and in revealing the details of the three-gluon mechanism of the OZI rule
breaking in $K\bar K$ channel.
\end{abstract}


\narrowtext

There are intentions to study the energy range of $e^+e^-$ annihilation
in the interval of the center-of-mass energy from $2E=1.5$ GeV up to
$m_{\rm J/\psi}$ using the collider VEPP-4M \cite{kurd1}. The BEPC
$e^+e^-$ collider team has also a plan to study some exclusive channels
in the energy range from 2 to 5 GeV \cite{xu}.
This raises the question of  comparison of the results of  existing
analysis of
the diagonal and transition form factors of light hadrons
in the energy range between 1 and 2 GeV \cite{ach97a,ach97b}
with the data now existing at the
$J/\psi$ mass. Here we perform this task, in order to
uncover possible surprises that might be revealed in future experiments.
We use the following formulas expressing the cross sections through
corresponding form factors.
If $h=\pi$ or $K$, then
\begin{equation}
\sigma(e^+e^-\to h^+h^-)=\frac{8\pi\alpha^2}{3s^{5/2}}|F_h(s)|^2p^3_h,
\end{equation}
where $p_h$ is the modulus of the 3-momentum of the hadron $h$ in the
center-of-mass system of $e^+e^-$ beams, whose total energy is $\sqrt{s}$.
The cross section for $\pi^+\pi^-$ is given by Eq. (2.1) of the paper
\cite{ach97a}. The cross section for $K^+K^-$ is given by
Eq. (2.1) of the paper \cite{ach97b}.
If the final state is $VP$, where $V,P=\omega,\pi^0(\rho^0,\eta)$, then
\begin{equation}
\sigma(e^+e^-\to VP)=\frac{4\pi\alpha^2}{3s^{3/2}}|F_{VP}(s)|^2p^3_{VP},
\end{equation}
where $p_{VP}$ is the modulus of the 3-momentum of the hadron $V$ (or $P$)
in the center-of-mass system of $e^+e^-$ beams.
The cross section for $VP$ final state is given by Eq. (2.1) of the paper
\cite{ach97a}.
If the final state is $2\pi^+2\pi^-$,  then
\begin{equation}
\sigma(e^+e^-\to 2\pi^+2\pi^- )=\frac{4\pi\alpha^2}{s^{3/2}}
|F_{\rho^0\pi^+\pi^-}(s)|^2W_{\pi^+\pi^-\pi^+\pi^-}(s),
\end{equation}
where cross section for the production of $2\pi^+2\pi^-$ is given by Eq. (2.8)
of Ref. \cite{ach97a}, and $W_{\pi^+\pi^-\pi^+\pi^-}(s)$ is given by Eq. (2.10)
of Ref. \cite{ach97a}.

The Okubo-Zweig-Iizuka (OZI) rule violating decays  of the $c\bar c$
quarkonia into the light hadrons are divided into two very different classes.
The isovector states $\pi^+\pi^-$,
$\omega\pi^0$, $\rho\eta$ and $\rho^0\pi^+\pi^-$ are produced
predominantly via the  one photon ($\gamma$)
intermediate state. The three-gluon ($ggg$) contribution which violates
the conservation of isospin should be suppressed. Indeed, in the
$\pi^+\pi^-$ channel, the ratio of the coupling constant due to  three
gluons to that due to one photon is estimated as
\begin{equation}
\frac{|a^{(ggg)}_\pi|}{|a^{(\gamma)}_\pi|}\sim{m_d-m_u\over Q}
\left({\alpha_s\over\pi}\right)^3\frac{f_{\rm J/\psi}}{4\pi\alpha
|F_\pi(m^2_{\rm J/\psi})|},
\label{ratio}
\end{equation}
where $\alpha=1/137$, $\alpha_s\simeq0.2$ is the  QCD coupling constant, and
$f_{\rm J/\psi}$ enters the expression for the leptonic width of the $J/\psi$
in a usual way:
\begin{equation}
\Gamma_{\rm J/\psi\to e^+e^-}={4\pi\alpha^2\over 3f^2_{\rm J/\psi}}
m_{\rm J/\psi}.
\label{leptw}
\end{equation}
Inserting $m_d-m_u\simeq3$
MeV, choosing conservatively $Q\sim m_\pi$, and taking the vector
dominance model (VDM) expression
\begin{equation}
F^{\rm(VDM)}_\pi(s)={m^2_\rho\over m^2_\rho-s}
\label{vdmff}
\end{equation}
for the pion form factor, one gets the figure of $10^{-2}$ for above ratio.
Similar estimate holds for other isovector channels cited above.
The amplitude with
$gg\gamma$ in intermediate state is also
expected to be suppressed \cite{milana}.
The production amplitude of the
isoscalar states includes the superposition of the
one photon and  $ggg$ amplitudes.
The production amplitude of strange mesons  includes the superposition of
both the isovector and isoscalar amplitudes.
First we will compare the data on the $J/\psi$ decays with the predictions of
the corresponding  VDM expression assuming the zero-width approximation
and then   to  more sophisticated
amplitudes which incorporate the complex mixing  of  mesons from the
ground state nonet with the heavier primed resonances \cite{ach97a,ach97b}.

Let us present our findings first  for the decay channels with the pair of
pseudoscalar mesons.
The modulus squared
of the pion form factor expressed through the ratio of partial
widths,
\begin{equation}
|F_\pi(m^2_{\rm J/\psi})|^2=4\frac{\Gamma(J/\psi\to\pi^+\pi^-)}
{\Gamma(J/\psi\to e^+e^-)},
\label{eq1}
\end{equation}
is $(11.9\pm1.5\pm0.9)\times10^{-3}$ \cite{balt85} [or slightly lower
figure of $(9.8\pm 1.5)\times10^{-3}$, according to the averaged value
of the $\pi^+\pi^-$ branching ratio found in \cite{pdg}]
and was already mentioned to be remarkably large
\cite{milana}. The VDM estimate according to Eq. (\ref{vdmff})
(see the dashed curved in Fig. \ref{figpi})
amounts to a figure of $4.3\times10^{-3}$.
In the case of  $\psi(2S)$
the pion form factor can be evaluated with the formula similar to
Eq. (\ref{eq1}) and gives, using the earlier DASP data \cite{brand},
the figure of $|F_\pi(m^2_{\psi(2S)})|^2=(36\pm23)\times10^{-3}$.
This is especially surprising since shows, guided by the
central figure,  the rise of the form factor with the energy increase, but,
certainly, experimental error is too large.
Using a more realistic amplitude which includes the $\rho^\prime_{1,2}$
resonances with the parameters obtained recently \cite{ach97a}, we plot
the corresponding curve with the dotted line in Fig. \ref{figpi}.
In this case the curve goes four times as low as compared to the experimental
value at the ${\rm J}/\psi$ mass. This is puzzling, since the one photon
contribution is the only way to explain the decay $\rm J/\psi\to\pi^+\pi^-$.

The above theoretical inconsistencies of the $\pi^+\pi^-$
channel strongly suggest that something new may happen at the energies
between 2 GeV and the mass of $J/\psi$, where the data are almost absent.
As an illustration, we add the resonance $\rho(2150)$ with the quantum
numbers $I^G(J^{PC})=1^+(1^{--})$ documented in the full listings of
Review of Particle Physics (RPP)
\cite{pdg}, ignoring, for nothing is better, the possible energy
dependence of its partial widths and the mixing with other $\rho$-like
resonances. Taking the mass $m_{\rho^\prime_3}=2010$ MeV, the width
$\Gamma_{\rho^\prime_3}=260$ MeV, the ratio of coupling constants
$g_{\rho^\prime_3\pi\pi}/f_{\rho^\prime_3}=0.08$, and slightly varying,
within the error bars, the parameters of the $\rho^\prime_{1,2}$
resonances found in Ref. \cite{ach97a}, one obtains the curve shown with
the solid line in Fig. \ref{figpi}. One can see that the knowledge of the
spectrum of still unknown isovector resonances (if any) above 2 GeV is
crucial for both the understanding of the behavior of the pion form factor
(and some other form factors, too, see below) and for establishing the limits
to applicability of the generalized VDM.

In general, the $K\bar K$ coupling of a C-odd quarkonium $J/\psi=c\bar c$
is represented in the form
\begin{equation}
g_{\rm J/\psi K\bar K}=a^{(ggg)}_K-{4\pi\alpha\over f_{\rm J/\psi}}
(\pm F^{(1)}_K+F^{(0)}_K),
\label{eq2}
\end{equation}
where $a^{(ggg)}_K$ being, in general, a complex number,
represents the pure isoscalar  contribution of the three gluons;
$F^{(I)}_K\equiv F^{(I)}_K(m^2_{\rm J/\psi})$
is the kaon electromagnetic form factor
with the given isospin $I=0,1$ taken at the $J/\psi$ mass \cite{fn1}.
The leptonic coupling constant $f_{\rm J/\psi}$ is expressed through leptonic
partial width by the expression  Eq. (\ref{leptw}).
The $K^+K^-$ and
$K_LK_S$ decay rates are distinguished by the sign of isovector
contribution, so that the ratio of $|F^{(1)}_K(m^2_{\rm J/\psi})|$ extracted
from the data \cite{balt85}, to the VDM estimate
\begin{equation}
|F^{(1)({\rm VDM})}_{K^+}(m^2_{\rm J/\psi})|={1\over2}
{m^2_\rho\over(m^2_{\rm J/\psi}-m^2_\rho)}=0.033
\label{vdmk}
\end{equation}
is found to be
2, 1, 2/3 for the relative phase of the $I=0$ and $I=1$ contributions
$\theta=62^\circ\mbox{, }22^\circ\mbox{, }0^\circ$, respectively. Note that
the latter case gives the lower bound to the isovector contribution. However,
the simple VDM amplitude fails to describe the data on the reaction
$e^+e^-\to K^+K^-$ in the energy range 2E=1.1$-$2 GeV; see Fig. \ref{figk}.
On the other hand, the isovector part of the  kaon form factor
extracted from the fit which includes the contributions of heavier
resonances $\rho^\prime_{1,2}$, $\omega^\prime_{1,2}$, and
$\varphi^\prime_{1,2}$ with the parameters found in Ref. \cite{ach97b},
can be matched with  isovector contribution extracted from the $J/\psi$
data, provided the
relative phase is $\theta=22^\circ$. Accidently, at the $J/\psi$ mass, the
absolute values of the isovector
kaon form factor in the simple VDM and in our fit
\cite{ach97b} turn out to be coincident.
In the meantime, the phase relations in the above
models are completely different. Specifically, one has
\begin{eqnarray}
F^{(0)}_K(m^2_{\rm J/\psi})&=&(6.5-6.3i)\times 10^{-3}, \nonumber\\
F^{(1)}_K(m^2_{\rm J/\psi})&=&(3.1-1.3i)\times 10^{-2},
\label{num}
\end{eqnarray}
with the set of parameters found in Ref. \cite{ach97b}. Note that the modulus
of the three-gluon coupling constant satisfies the relation
$|a^{(ggg)}_K|\geq 0.9|g_0|$ regardless the relative phase between the
three-gluon contribution and isoscalar part of the one photon one.
Here the numerical factor of 0.9
comes from  the numerical value of the isoscalar kaon form factor given in
Eq. (\ref{num}), and $g_0$ is the coupling constant
of  $J/\psi$ to $K\bar K$  in the $I=0$ state. Since one can hardly
imagine the mechanism of  enhancement of the isoscalar form factor
by an order of magnitude in comparison with that given in Eq. (\ref{num}),
we see that the greater
part of the isoscalar coupling constant is due to the three-gluon
contribution. There are no reasons
to neglect the latter  and attribute all the $K\bar K$ branching
ratio of the $J/\psi$ solely to the one photon mechanism, as it was assumed
in Ref. \cite{balt85}.

Now turn to the vector and  pseudoscalar final states
\cite{balt85b,cof88,jouss}.
The ratio of the absolute values of the $\omega\pi^0$ form factors is
expressed through the measured branching ratios as \cite{balt85b,cof88}
\begin{eqnarray}
\frac{|F_{\omega\pi^0}(m^2_{\rm J/\psi})|}
{|F_{\omega\pi^0}(0)|}&=&\left[{\alpha\over3}\left({q_{\gamma\pi^0}\over
q_{\omega\pi^0}}\right)^3\right.         \nonumber\\
& &\left.\times\frac{m_{\rm J/\psi}\Gamma(J/\psi\to\omega\pi^0)}
{\Gamma(\omega\to\gamma\pi^0)\Gamma(J/\psi\to\mu^+\mu^-)}\right]^{1/2}.
\label{eq3}
\end{eqnarray}
The VDM evaluation of the above ratio gives a figure of
0.0659 which is by a factor
of two greater than the experimentally measured figure of $0.0335\pm0.0059$
\cite{cof88}. On the other hand, the inclusion of the $\rho^\prime_{1,2}$
resonances \cite{ach97a} interfering destructively with the $\rho(770)$
tail at energies above 2 GeV results in the calculated figure to be
twice as low
as experimentally measured. See the curve in Fig. \ref{figompi}. The
result of the calculation of an analogous ratio for the $\rho\eta$ final state
is shown in Fig. \ref{figrhe}.
Finally, the form factor of the $\rho^0\pi^+\pi^-$
final state which enters the partial width of the $J/\psi$ as
\begin{eqnarray}
\Gamma(J/\psi\to\rho^0\pi^+\pi^-)&=&12\pi\alpha\frac{\Gamma(J/\psi
\to\mu^+\mu^-)}{m_{\rm J/\psi}}      \nonumber\\
& &\times\left|F_{\rho^0\pi^+\pi^-}
(m_{\rm J/\psi}^2)\right|^2         \nonumber\\
& &\times W_{\pi^+\pi^-\pi^+\pi^-},
(m_{\rm J/\psi}^2)
\label{eq4}
\end{eqnarray}
where $W_{\pi^+\pi^-\pi^+\pi^-}$ is the phase space volume of the
$2\pi^+2\pi^-$ state given in \cite{ach97a},
analogously for the $\psi(2S)$, is plotted in Fig. \ref{fig4pi}. The VDM
estimate in this case is
\begin{equation}
F_{\rho^0\pi^+\pi^-}(s)={2g_{\rho\pi\pi}m^2_\rho\over m^2_\rho-s},
\label{vdm4pi}
\end{equation}
where the relation among the coupling constants $g_{\rho^0\rho^0\pi^+\pi^-}
=2g^2_{\rho\pi\pi}$ resulting from the  vector current conservation
is taken into account, together with the neglect of the bremstrahlung-type
diagrams. See Ref. \cite{ach97a} for some details of approximations made
for the $\rho\rho\pi^+\pi^-$ coupling.
Both curves go far below the $J/\psi$ and $\psi(2S)$ data. Note that the
$\rho^0\pi^+\pi^-$ contribution was not isolated in the total
$\pi^+\pi^-\pi^+\pi^-$ data sample at the $J/\psi$ mass \cite{jean}.
However, such an isolation was implemented at the $\psi(2S)$ mass, and
the $\rho^0\pi^+\pi^-$ contribution was found to be $93\%$ \cite{tanen} of
the total  number of $\pi^+\pi^-\pi^+\pi^-$ events.
Since one cannot foresee any reason
why the situation, in this respect,
at the $J/\psi$ could differ from the $\psi(2S)$, we
simply insert $B(J/\psi\to\pi^+\pi^-\pi^+\pi^-)$ in place of
$B(J/\psi\to\rho^0\pi^+\pi^-)$, in order to find the
$\rho^0\pi^+\pi^-$ transition form factor at the $J/\psi$ mass.

Since in almost all cases the curves in Fig. \ref{figpi}$-$\ref{fig4pi} go
well below the $J/\psi$ data points, one can see that
some isovector resonance structures with the masses
above 2 GeV interfering strongly with those already included are likely
to be present. The example of the $\pi^+\pi^-$ channel shows that the fit 
of the data with the  $J/\psi$ data point included is 
improved with the $\rho^\prime_3$ resonance being taken into account
\cite{fn2}.
Their isoscalar partners are also rather probable. They could manifest
themselves in the channels of $e^+e^-$ annihilation into $\omega\eta$,
$\omega\eta^\prime$, $\rho\pi$, $\omega\pi^+\pi^-$ etc and in the
decay channels which include strange particles. All this suggests that
the energy region above 2 GeV of
$e^+e^-$ annihilation is interesting from the point of view of elucidating
the spectrum of states with the  masses in this range
and for establishing
the detailed form (modulus and  phase) of the three-gluon coupling with
different states including its dependence on energy.
To gain an impression of what the typical cross section magnitudes
might be, we give the calculated figures at the energy $\sqrt{s}=2.5$
GeV. In the case of the final states $\pi^+\pi^-$, $K^+K^-$, $\omega\pi^0$,
$\rho^0\eta (\pi^+\pi^-\eta)$, and $\pi^+\pi^-\pi^+\pi^-$ they are,
respectively, 0.03, 0.02, 0.04, 0.04, and 0.6 nanobarns.

\acknowledgments

The present work was supported in part by the grant INTAS-94-3986.

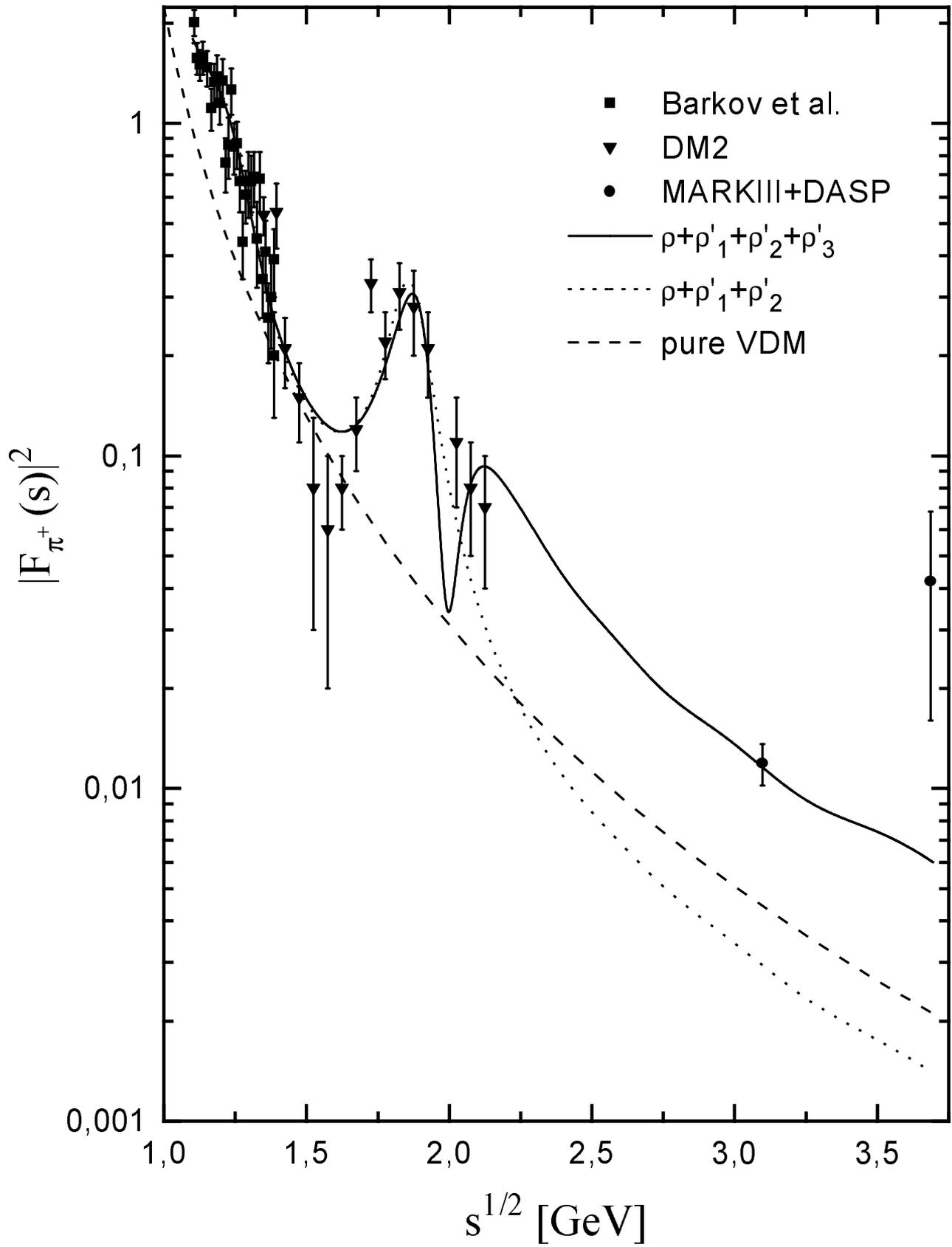
\begin{figure}
\caption{The pion form factor. The data are from Barkov et al.
\protect\cite{barkov85}, DM2 \protect\cite{dm2pi}, MARKIII
\protect\cite{balt85}, DASP \protect\cite{brand}.\label{figpi}}
\end{figure}
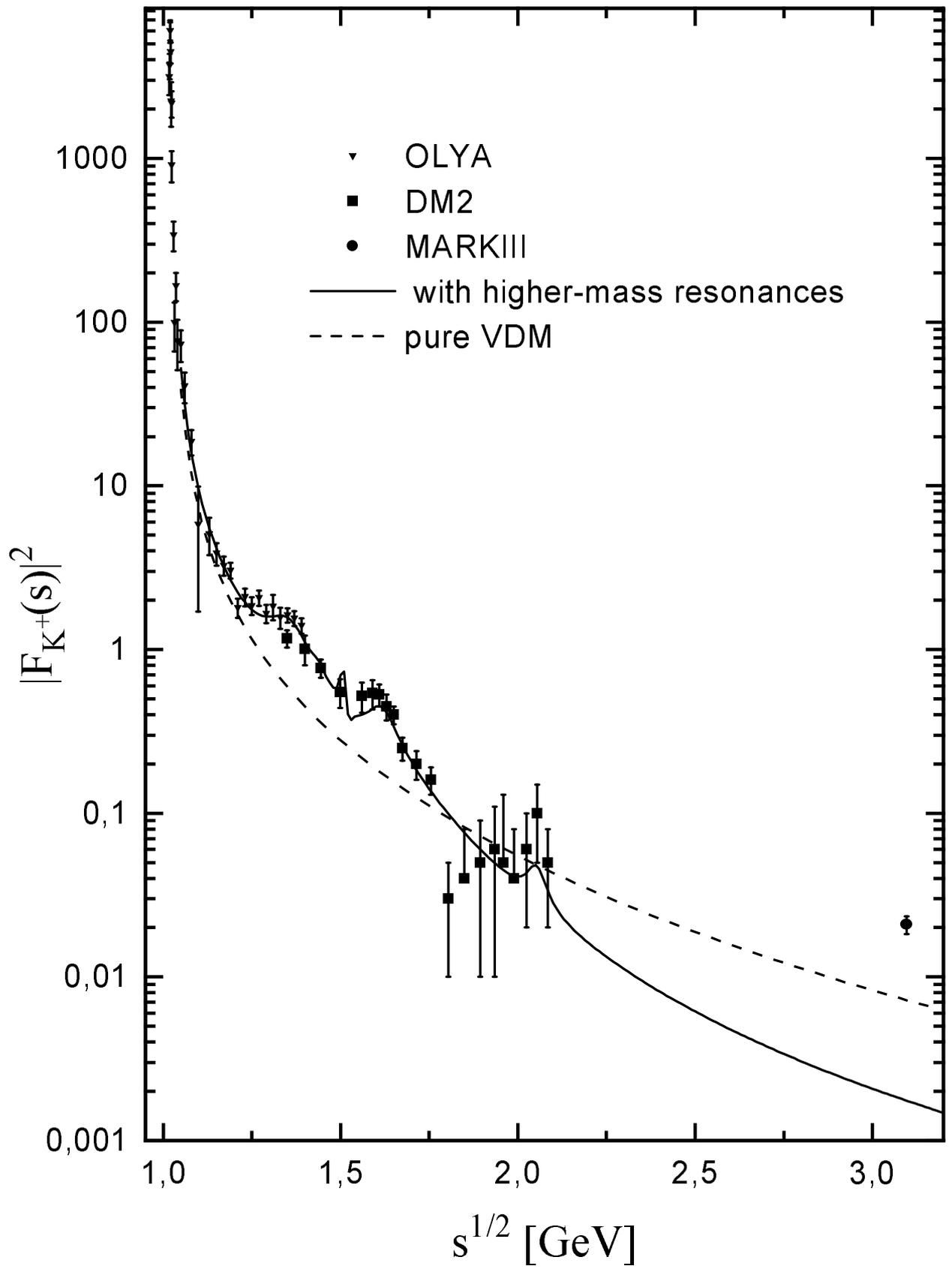
\begin{figure}
\caption{The charged kaon form factor. The experimental point at the
$J/\psi$ mass is given by the authors of
Ref. \protect\cite{balt85}
upon neglecting the three-gluon contribution. The data are from
OLYA \protect\cite{olya}, DM2 \protect\cite{dm2_kk},
MARKIII \protect\cite{balt85}. The solid curve is drawn upon taking into
account the contributions of higher mass resonances $(\rho^\prime_1+
\omega^\prime_1+\varphi^\prime_1)+(\rho^\prime_2+
\omega^\prime_2+\varphi^\prime_2)$, with the parameters found in
\protect\cite{ach97a,ach97b}.\label{figk}}
\end{figure}
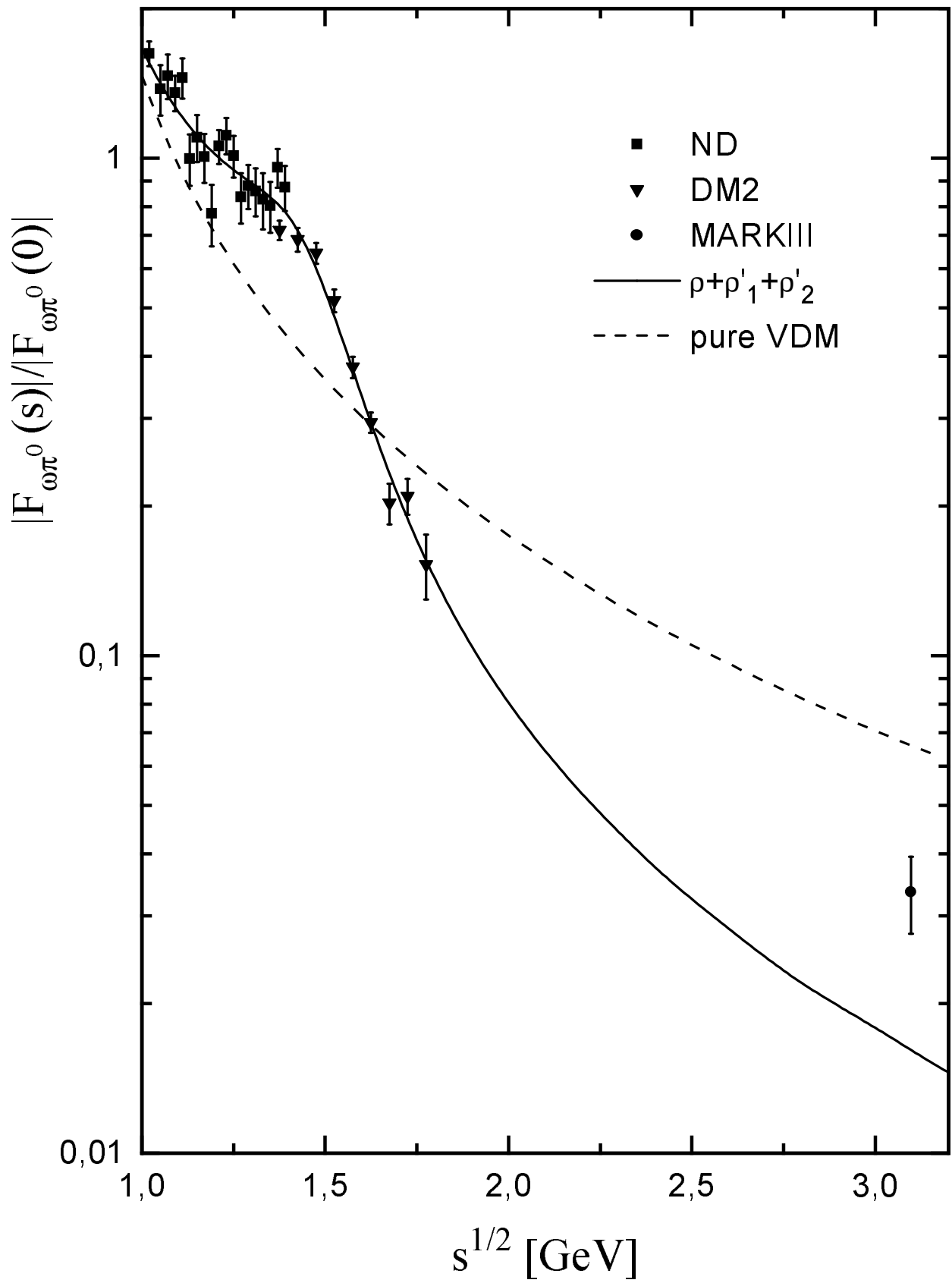
\begin{figure}
\caption{The $\omega\pi^0$ form factor. The data are recalculated from the
cross section data of ND \protect\cite{dol91} and DM2 \protect\cite{stan91}.
\label{figompi}}
\end{figure}
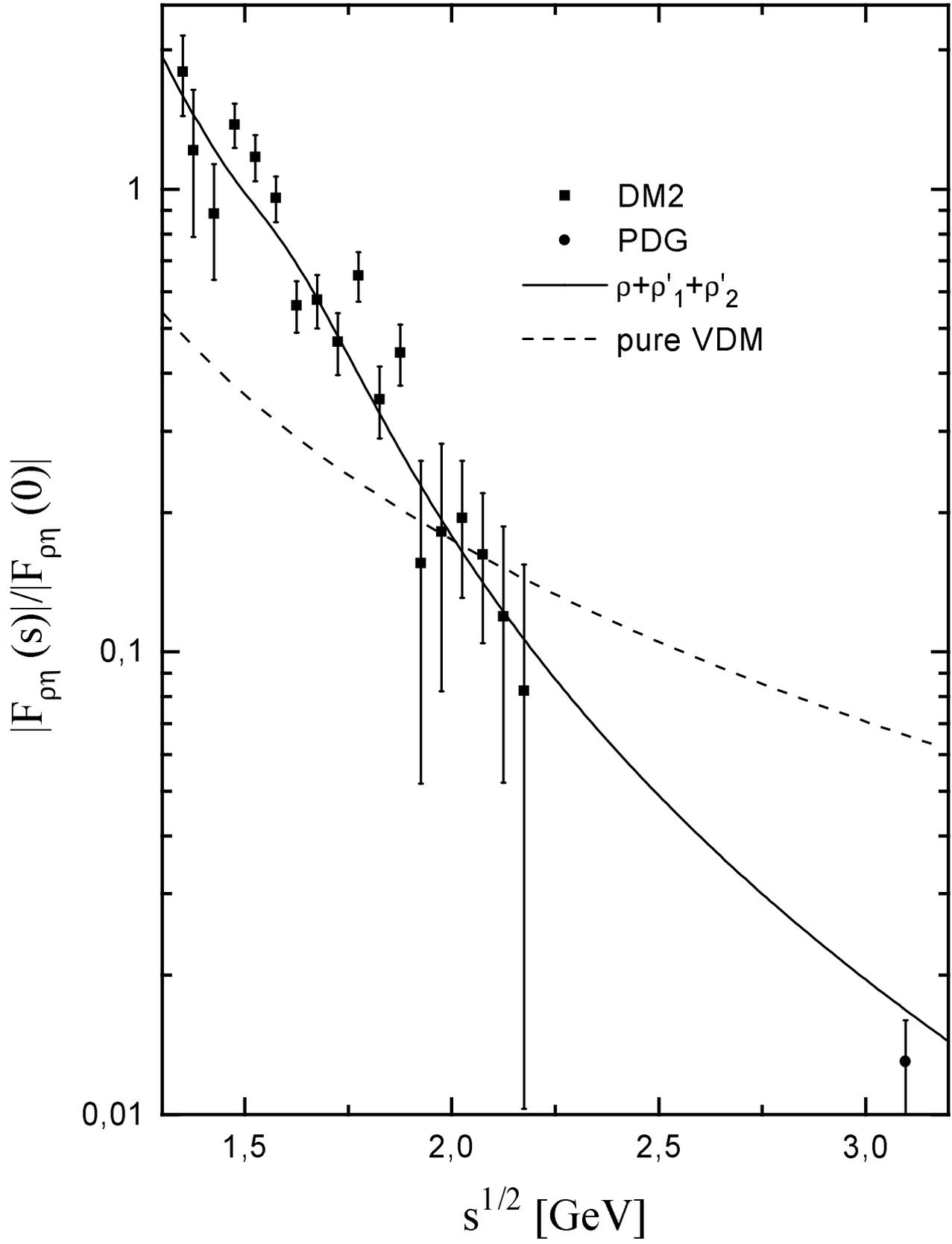
\begin{figure}
\caption{The $\rho\eta$ form factor. The DM2 data are recalculated from the
cross section data of \protect\cite{ant},
PDG \protect\cite{pdg}. \label{figrhe}}
\end{figure}
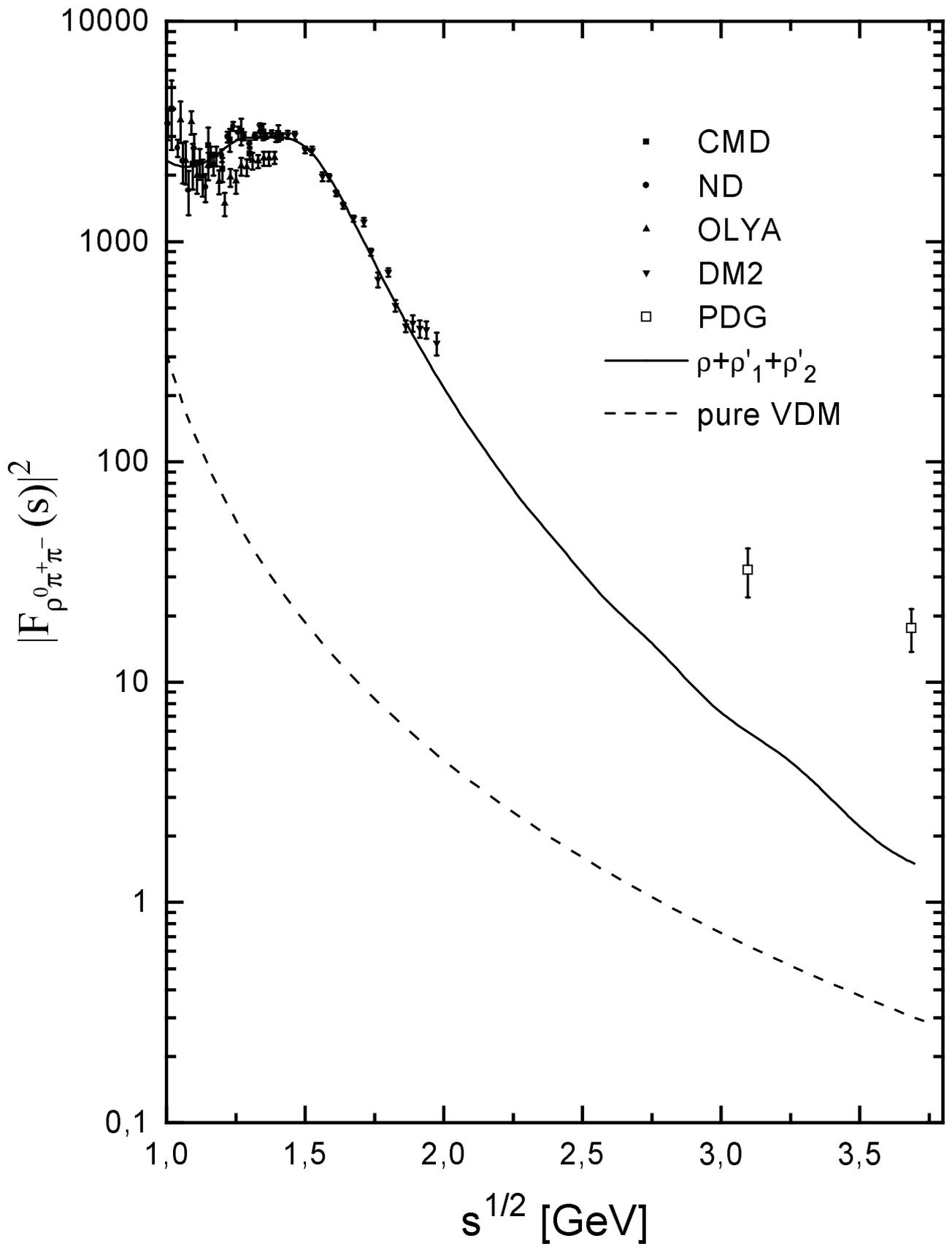
\begin{figure}
\caption{The $\rho\pi^+\pi^-$ form factor. The data are recalculated
from the cross section data of CMD \protect\cite{bar88}, ND
\protect\cite{dol91},
OLYA \protect\cite{kurd}, DM2 \protect\cite{stan91}; MARKI
\protect\cite{jean,tanen}.\label{fig4pi}}
\end{figure}

\begin{references}
\bibitem{kurd1}
L.~M.~Kurdadze, private communication.
\bibitem{xu}
Xu Gou-fa, private communication.
\bibitem{ach97a}
N.~N.~Achasov and A.~A.~Kozhevnikov,  Phys. Rev. D{\bf55}, 2663 (1997);
hep-ph/9609216.
\bibitem{ach97b}
N.~N.~Achasov and A.~A.~Kozhevnikov, Phys. Rev. D{\bf57}, 4334 (1998);
hep-ph/9703397.
\bibitem{milana}
J.~Milana, S.~Nussinov and M.~G.~Olsson, Phys. Rev. Lett., {\bf71},
2533 (1993).
\bibitem{balt85}
R.~M.~Baltrusaitis {\it et al.}, Phys. Rev. D{\bf32}, 566 (1985).
\bibitem{pdg}
R.~M.~Barnett {\it et al.} (Particle Data Group), Phys. Rev.
D{\bf 54}, 1 (1996).
\bibitem{brand}
R.~Brandelik {\it et al.}, Z. Phys. C{\bf1}, 233 (1979).
\bibitem{fn1}
The arguments in favor of neglecting the one-photon coupling to the neutral 
kaon pair put forward in Ref. \cite{milana} would be fulfilled only in the 
case of exact degeneracy of the $\rho(770)$, $\omega(782)$, and 
$\varphi(1020)$ masses. In reality, however,
one obtains the ratio 
$|F_{K^+}(m^2_{J/\psi})|/|F_{K^0}(m^2_{J/\psi})|$ equal to
4.5 and 1.7 in the simple VDM and in the VDM with inclusion of heavier
resonances \cite{ach97a,ach97b}, respectively.
\bibitem{balt85b}
R.~M.~Baltrusaitis {\it et al.}, Phys. Rev. D{\bf32}, 2883 (1985).
\bibitem{cof88}
D.~Coffman {\it et al.}, Phys. Rev. D{\bf38}, 2695 (1988).
\bibitem{jouss}
J.~Jousset {\it et al.}, Phys. Rev. D{\bf41}, 1389 (1990).
\bibitem{jean}
B.~Jean-Marie {\it et al.} Phys. Rev. Lett., {\bf 36}, 291 (1976).
\bibitem{tanen}
W.~Tanenbaum {\it et al.} Phys. Rev. D{\bf 17}, 1731 (1978).
\bibitem{fn2}
Because of the lack of data above 2 GeV, the coupling constants of the
$\rho^\prime_3$ resonance with specific channels are unknown, so we do not
include it in other channels except the $\pi^+\pi^-$ one.
The procedure of extracting the $\rho^\prime_3$ couplings,
with the only $J/\psi$ data points at hand, would appear
to be speculative, since the $\chi^2$ criterion  determined in that case
essentially by the low energy data points, would not put any serious 
restrictions on magnitudes of these couplings.
\bibitem{barkov85}
L.~M.~Barkov {\it et al.}, Nucl. Phys. {\bf B256}, 365 (1985).
\bibitem{dm2pi}
DM2 Collaboration, D.~Bisello {\it et al.}, Phys. Lett. B{\bf220}, 321
(1989).
\bibitem{olya}
P.~M.~Ivanov {\it et al.} Phys. Lett {\bf 107B}, 297 (1981).
\bibitem{dm2_kk}
DM2 Collaboration, D.~Bisello {\it et al.} Z. Phys. C{\bf39}, 13 (1988).
\bibitem{dol91}
S.~I.~Dolinsky {\it et al.}, Phys. Rep. {\bf202}, 99 (1991).
\bibitem{stan91}
L.~Stanco, in {\it Hadron 91}, Proceedings of the International Conference
on Hadron Spectroscopy, College Park, Maryland, edited by S.~Oneda
and D.~C.~Peaslee (World Scientific, Singapore, 1992) p. 84.
\bibitem{ant}
DM2 Collaboration, A.~Antonelli {\it et al.}, Phys. Lett. B{\bf212},
133 (1988).
\bibitem{bar88}
L.~M.~Barkov {\it et al.} Yad. Fiz. {\bf47}, 393 (1988).
\bibitem{kurd}
L.~M.~Kurdadze {\it et al.}, Pis'ma Zh. Eksp. Teor. Fiz. {\bf47}, 432 (1988).

\end{references}
\end{document}